\documentclass[aps,prl,twocolumn,groupedaddress]{revtex4-1}

\usepackage{graphicx}
\usepackage{amsmath}

\begin{document}
\title{Method for Cooling Nanostructures to Microkelvin Temperatures}


\author{A. C. Clark}
\email{Both authors contributed equally to this work.}
\author{K. K. Schwarzw\"alder}
\email{Both authors contributed equally to this work.}
\author{T. Bandi}
\author{D. Maradan}
\author{D. M. Zumb\"uhl}
\email[]{Dominik.Zumbuhl@unibas.ch}
\affiliation{Department of Physics, University of Basel, Klingelbergstrasse 82, CH-4056 Basel, Switzerland}

\date{\today}

\begin{abstract}
We propose a new scheme aimed at cooling nanostructures to microkelvin temperatures, based on the well established technique of adiabatic nuclear
demagnetization: we attach each device measurement lead to an individual nuclear refrigerator, allowing efficient thermal contact to a microkelvin
bath. On a prototype consisting of a parallel network of nuclear refrigerators, temperatures of $\sim 1\,$mK simultaneously on ten measurement leads
have been reached upon demagnetization, thus completing the first steps toward ultracold nanostructures.
\end{abstract}


\maketitle 

The ability to reach low millikelvin or even microkelvin temperatures in nanoscale samples would open up the possibility to discover new physics in a
variety of systems. For example, an intriguing nuclear spin ferromagnetic phase transition in a GaAs interacting 2D electron gas (2DEG) has been
predicted to occur around $\sim 1\, $mK \cite{loss1,*loss2} at $B = 0$, constituting a novel type of correlated state. Nuclear spin fluctuations
would be fully suppressed in this ferromagnetic phase, eliminating the main source of decoherence for GaAs spin qubits \cite{SpinQubitReviewRMP}.
Further, full thermodynamic nuclear polarization is possible at temperatures $T \lesssim 1\,$mK in an external magnetic field of $B \sim 10\,$T
\cite{chesi}. Other systems benefiting from ultralow temperatures include fractional quantum hall states with small gaps \cite{pan,*xia}, in
particular the $\nu=5/2$ state \cite{willet}, which is currently considered for topological quantum computation \cite{kitaev,nayak}.

The majority of quantum transport experiments to date, such as those in GaAs 2DEGs or any other nanoelectronic devices on insulating substrates, have
been carried out at electron temperatures $T_e$ significantly greater than that of the host $^3$He-$^4$He dilution refrigerator (DR). Since only
metals provide significant thermal conduction at temperatures well below 1 K \cite{lounasmaa,pobell}, nanostructures are thermally connected to and
cooled by the DR primarily through their electrical leads. Since these leads need to be electrically isolated, some insulator will still inhibit
efficient cooling. The main challenges for cooling such samples below 1 mK include overcoming poor thermal coupling between electrons in the leads
and the refrigerator \cite{chung}, providing sufficient attenuation of high frequency radiation, and reducing low frequency interference such as
ground loops. To our knowledge, the minimum temperature reported is 4 mK, with sintered silver heat exchangers attached to sample wires in a $^3$He
cell \cite{pan,*xia}. Similarly, Pomeranchuck cooling \cite{pobell} with sinters on each sample wire could reach temperatures $\sim$ 1 mK.

\begin{figure}[t]
\scalebox{1}[1] {\includegraphics[height=3in]{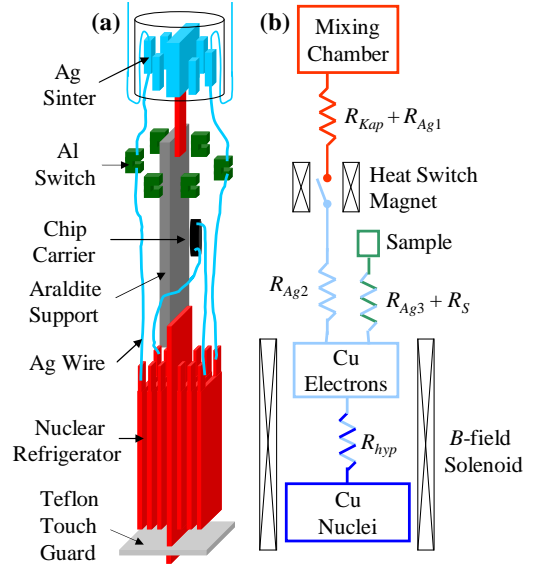}}\vspace{-2mm}\caption{\label{fig1} (Color) (a) Schematic of parallel nuclear refrigerator
network. Only six NRs are shown for simplicity. (b) Cooling scheme, with different colors denoting potentially different temperatures under steady
state conditions.\vspace{-4mm}}
\end{figure}

In this Letter, we present a new method intended to cool nanostructures into the microkelvin regime. We propose to adapt the very well established
technique of nuclear adiabatic demagnetization to the specific needs of nanoscale samples: every sample wire passes through its own, separate nuclear
refrigerator (NR) (see Fig.~\ref{fig1}), ensuring excellent thermal contact even at microkelvin temperatures between the sample and the NR while
keeping all wires electrically isolated from one another, as required for measurements. With this method, nanostructures can in principle be cooled
to less than 100 $\mu$K, which would be a reduction in temperature by more than two orders of magnitude compared with common $T_e \gtrsim 10\,$mK.
Further, we have designed, built, and tested a prototype refrigerator that is based on this proposal. We present evidence for achieving $\sim 1\,$mK
in ten NRs simultaneously, thus completing the first steps towards microkelvin nanostructures.

Nuclear adiabatic demagnetization is the most widely used technique available today for ultralow temperature experiments in condensed matter
\cite{pickett,lounasmaa,pobell}. The lowest temperatures reported are $\sim 1\,\mu$K \cite{wendler} for electrons in Pt and $\sim 300\,$pK for
nuclear spins in Rhodium \cite{knuuttila, hakonen} -- among the lowest temperatures achieved in any laboratory. It is a single shot method consisting
of three steps. First, a suitable metal with nonzero nuclear spin (the NR) is precooled with a DR in a large magnetic field $B_i$ to a
temperature $T_i\sim 10\,$mK, generating as large a thermodynamic nuclear spin polarization as possible. Second, thermal contact between the DR and
NR is cut off by a superconducting heat switch and the field is adiabatically reduced by a large factor, e.g. $x=B_i/B_f\sim100$. Ideally (perfect
adiabaticity), the nuclei are cooled by the same factor such that $T_f=T_i/x$. Finally, experiments are performed at microkelvin temperatures for a
finite time, typically days or even weeks. The heat leaking into the system plays an important role since it increases $T_e$ in the NRs above
the nuclear spin temperature and is absorbed by the nuclei until the polarization is lost and the NRs heat up to or above DR temperatures.

We propose to cool nanostructures to microkelvin temperatures using a parallel network of NRs. Each NR constitutes part of the electrical connection
from room temperature down to the sample. Semiconductors and insulators, commonly used in nanosamples, are not practical as NRs since it is
difficult to sufficiently precool their nuclei. Still, one might consider as NRs devices with large conducting regions containing nuclear spins, such
as GaAs 2DEGs with a highly doped, metallic back gate (or similar). However, their nuclear heat capacity would be drained all too
quickly given realistic heat leaks. Therefore, our strategy is to incorporate the most widely used material for NRs: Cu, an excellent conductor with
nuclear spin 3/2. In this system the nuclear hyperfine interaction couples the nuclei at temperature $T_f$ to the electrons at temperature $T_e$ with
a characteristic nuclear spin relaxation time $T_1$ that obeys the Korringa law, $T_e T_1 = \kappa\approx1\,$K$\,$s \cite{lounasmaa,pobell}. The
effective thermal equilibration time is reduced from $T_1$ by the very large ratio of nuclear and electronic heat capacities, resulting in strong,
fast coupling even at $T_e < 100\,\mu$K \cite{george}. However, conducting sample sections may be thermally isolated from other degrees of
freedom at low enough temperatures.

We now turn to the discussion of the prototype NR network (see Fig.~\ref{fig1}). Each NR consists of a Cu plate situated in the center of a
demagnetizing field, connected with high conductivity wires on one side to a home built chip carrier made from 2850 FT Stycast epoxy \cite{rich}, and
on the other side through a heat switch \cite{switch} to the mixing chamber (MC) of a DR. Twelve parallel NRs are tied to a sacrificial NR with
dental floss, using small teflon spacers ensuring electrical (and thermal) isolation, giving a total of thirteen NRs. The sacrificial Cu piece is
glued into an araldite \footnote{Supplied by Huntsman Advanced Materials GmbH.} beam extending from the MC and holds a teflon touch guard in place,
which keeps the NRs from contacting the radiation shield.

To ensure proper operation of the NRs, we note some important details. Each measurement wire begins with 1.6 m of Thermocoax acting as a microwave
filter between room temperature and the MC cold finger. It then passes through a silver epoxy microwave filter \cite{us} and is transferred to a bare
Ag wire that is fed directly into the plastic MC. For efficient thermalization of the Cu during precooling the thermal resistance between the NR and MC must be minimized. We therefore use annealed, high purity Ag wire with 1.27 mm diameter and residual resistivity ratio $\geq$ 1500, which are spot welded to the Cu pieces and, in the MC, sintered to Ag nanoparticles (yielding surface areas of $\sim3\,$m$^2$ per wire used to overcome the Kapitza resistance $R_{Kap}$). The heat switches are ``C''-shaped pieces of annealed, high purity Al fused to the Ag wires on both ends. The small critical field of 10 mT allows easy switching with a home built magnet. The ratio of thermal conductivities in the closed state (Al normal) to the open state (Al superconducting) exceeds $10^4$ below 20 mK. Al pieces are placed carefully to minimize differences in the stray $B$-field from the solenoid, adding additional complexity for a network of NRs.

\begin{figure}[t]
\scalebox{1}[1] {\includegraphics[width=3.3in]{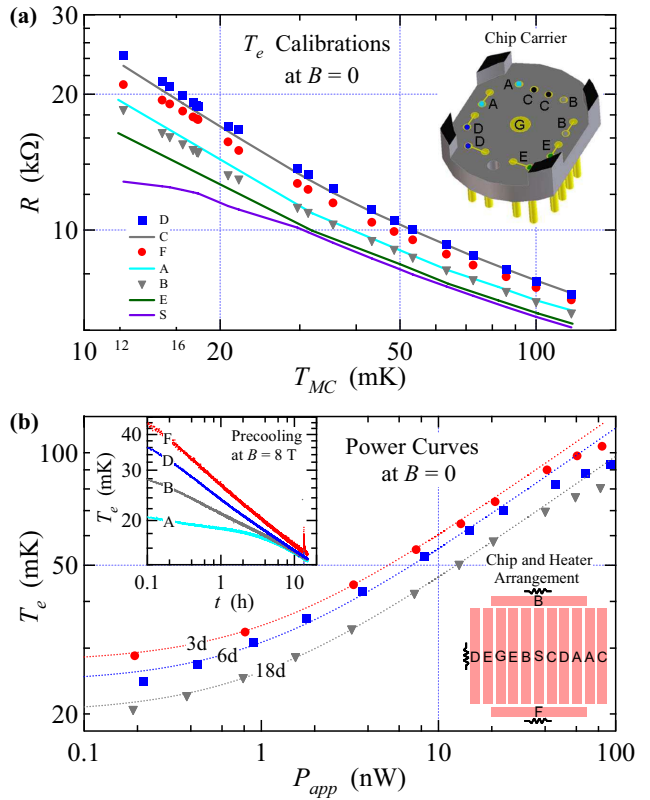}}\vspace{-3mm}\caption{\label{fig2} (Color) (a) Resistance of NR RuO$_2$ chips as a function
of $T_{MC}$. Inset: chip carrier. (b) Power curves with Al switches open, measured 3, 6, and 18 days after cooling down, shown for chips F (circles),
D (squares), and B (triangles), respectively. Dashed lines are theory (see text). Left inset: $T_e$ vs. time during precooling ($B=8\,$T). Right
inset: arrangement of Cu plates in the NR stack.\vspace{-4mm}}
\end{figure}

The entire stage is removable at a plastic cone seal at the MC, allowing samples to be directly wire bonded to polished Ag wires. Probably the
weakest thermal/electrical link between the device and the NR occurs at the Schottky barriers of the metal-semiconductor contacts, integrated on
chip. In steady state, parasitic heat leaking into the device will equal the heat leaving it through its thermal links to the NRs, setting the lowest
achievable temperature. Metallic nanostructures will benefit from comparatively higher conductivity metal-metal contacts.

Characterization of the NRs has been carried out by monitoring $T_e$ of the various Cu plates. Five RuO$_2$ chip resistors, labeled A-E, were mounted
on the chip carrier (see Fig.~\ref{fig2}(a) inset) and electrically connected \footnote{The (superconducting) tin on the pads of the RuO$_2$ chips
was removed. Contact was made using Ag epoxy.} to ten of the thirteen NRs, with each chip using a pair of NRs as its leads. The resistance reading in
these cases reflects an ``average'' temperature of each pair. Two more chips, F and S, were directly mounted onto individual NRs (S on the
sacrificial Cu plate), with the second contact of each electrically connected to -- but thermally isolated from -- the outside world by a bare NbTi
superconducting wire. The final plate was left unmonitored, serving as electrical ground (G) for chip capacitors across A and B. It is well known
that RuO$_2$ thermometers can suffer from rather long time constants \footnote{The total thermal equilibration times for the chips are 20 min at 30
mK, 120 min at 15 mK, and $>4\,$h at 5 mK.} and saturate below 10 mK. However, in the demagnetized state we can extrapolate the NR temperatures below
10 mK based on warm up curves, as will be described below.

We first calibrate the RuO$_2$ thermometers between 12 and 120 mK. Figure~\ref{fig2}(a) shows the resistance of the seven RuO$_2$ chips at $B=0$ as a
function of mixing chamber temperature $T_{MC}$ with Al switches closed. $T_{MC}$ was measured by a cerium magnesium nitrate (CMN) thermometer bolted
to the MC cold finger. Before measuring each data point, an appropriate amount of time for thermalization was allowed. There is no apparent
saturation down to $T_{MC}=12\, $mK for thermometers A-F, which all exhibit qualitatively the same temperature dependence. Moreover, on two separate
cooldowns a second CMN was mounted directly onto one of the NRs (first A, then F), verifying that $T_e$ measured by the RuO$_2$ chips is indeed equal
to $T_{MC}$ \footnote{The NR was able to cool the CMN to $\leq3\,$mK. However, a heat leak of $>20\,$nW was detected from the thermometer, severely
limiting the NR performance.}. We therefore use the data in Fig.~\ref{fig2}(a) as electron temperature calibrations for NRs A-F. The sacrificial
plate thermometer S displays some saturation for $T_{MC}\leq30\,$mK, presumably due to some heat leak. Upon ramping the magnet to $B=8\,$T, the
massive nuclear heat of magnetization significantly elevates the NR temperatures (varying somewhat for individual plates depending on stray $B$-field
conditions at the corresponding heat switches), but they cool within 15 h to $T_i = T_e\leq 15\,$mK (see Fig.~\ref{fig2}(b) left inset).

Next, we measure the parasitic heat leak to the NRs. Figure~\ref{fig2}(b) shows the NR temperature $T_e$ as a function of applied power at $B=0$ and
with the Al switches open. The heat flowing into the Cu is ultimately drained away by the MC, with the superconducting Al piece as the primary
impedance. Its thermal conductance is dominated at low temperature by phonon-dislocation scattering processes, obeying the relation \cite{pobell,
Alswitch}, $P_{app}=nA(T_e^3-T_{MC}^3)-P_0$, where $P_{app}$ is the applied power, $n=0.57\,$mol of Cu, $A$ is a prefactor, and $P_0$ is the
intrinsic heat leak to the NR. For $T_e>70\,$mK, parallel channels of heat flow become accessible \footnote{Heat conduction by free electrons becomes
larger than phonons near 70 mK. Also, there will be some heat flow through the teflon spacers between NRs at high $T_e$.}. Fits (dashed curves) are
in very good agreement with the data and allow us to extract $P_0$, which improved over time as indicated by the decrease in $T_e$ (true for \emph{all} chips A-F) as $P_{app} \rightarrow$ 0 in the power curves obtained 3, 6, and 18 days after cooling down (top to bottom curves). We conclude that the typical intrinsic heat leak to the NR stage at $B=0$ is $P_0/n \lesssim 1\,$nW$\,$mol$^{-1}$, sufficiently low but clearly above the state of the art value of $<5$ pW$\,$mol$^{-1}$ \cite{george}. Heat leaks measured at $B=1\,$T are also at the nanowatt level and similar for all NRs.

\begin{figure}[t]
\scalebox{1}[1] {\includegraphics[width=3.2in]{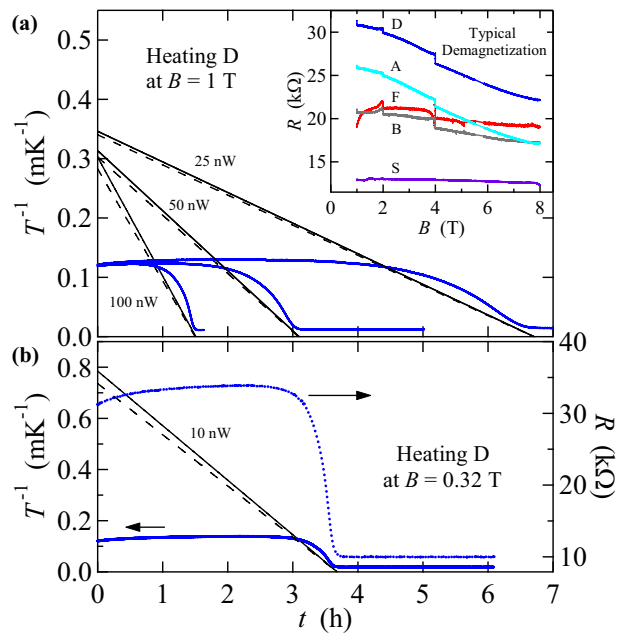}}\vspace{-3mm}\caption{\label{fig3} (Color) (a) Systematic heating tests for chip D after
demagnetization. Measured $T_e^{-1}$ (blue curves) and theory for $T_f^{-1}$ (solid lines) and $T_e^{-1}$ (dashed lines) vs. $t$ for the applied
powers indicated. Inset: cooling of chips during demagnetization. $R$ of thermometer D is plotted against the right axis for comparison in
(b).\vspace{-4mm}}
\end{figure}

Given a heat leak sufficiently low for nuclear cooling, we now evaluate the demagnetization process itself, starting from $T_i=15\,$mK and
$B_i=8\,$T. The inset of Fig.~\ref{fig3} shows the resistance of several chips during a series of ramps from 8 T $\rightarrow$ 1 T with open heat
switches. The field is decreased linearly in time using three sequential ramps at 1 T$\,$h$^{-1}$ from 8 T $\rightarrow$ 4 T and 4 T $\rightarrow$ 2
T, and at 0.5 T$\,$h$^{-1}$ from 2 T $\rightarrow$ $B_f=1\,$T. $R$ values increase upon demagnetization, clearly indicating cooling. They continue to
increase (at fixed field values of 4 and 2 T) between each field ramp and in fact do so more quickly, reflecting both a thermal lag between the chips
and Cu plates as well as a sensitivity to ramp rate. If enough time is allowed after reaching 1 T, $R$ increases further by 5 to 10 k$\Omega$ for
chips A-E (not shown). Chip F warms up near the end of the demagnetization, while S is nearly constant. It is presently unclear how the stored energy
in the Cu nuclei is drained from F, but its performance improves for lower precooling temperatures. For S, the lack of cooling \emph{and} warming
suggest that thermal contact 
to the environment remains significant.

We extract $T_f$ and $T_e \,(\gtrsim T_f$) of the NRs reached after demagnetizing to $B_f$ by recording the time $t$ necessary for a Cu plate to
``completely'' ($T_e^{-1}\rightarrow 0$) warm up under an applied power, using $t=n\Lambda B_f^2 / (P_{app} T_{f})$ and $T_e$=$T_f(1+\kappa P_{app}
/n\Lambda B_f^2)$, where $\Lambda$ is the nuclear Curie constant for Cu \cite{pobell,george}. In Fig.~\ref{fig3}(a) we plot time traces of $T_e^{-1}$
for chip D (not calibrated above 0.08 mK$^{-1}$) for $P_{app}=$ 25, 50, and 100 nW. The other thermometers give consistent results, except for F,
which appears to heat up during demagnetization. Due to poor internal thermalization $T_e^{-1}$ increases for the first several hours despite the
influx of heat, only showing significant signs of warming once the Cu is hotter than $\sim$10 mK. Since the temperature gradient between the chip and
NR will vanish at $T_f^{-1} = T_e^{-1} = 0$, we fix this point of the theoretical $T^{-1}$ curves (solid and dashed lines) and extrapolate back to $t
= 0$. As expected, larger $P_{app}$ results in faster warm up times. With this, we obtain $T_f(0)$ = 3.0$\pm$0.3 mK for all three $P_{app}$,
demonstrating the reliability of achieving a particular minimum temperature for a set demagnetization parameters. The uncertainty in $T_f(0)$ is
dominated by the inhomogeneity of $B_f$. We note that in the temperature range explored here, $T_e(0)\approx T_f$(0) before the power is turned on
since $P_0<<P_{app}$.

The final test from the present work (see Fig.~\ref{fig3}(b)) is a demagnetization from 8 T $\rightarrow$ 0.32 T starting at 13.3 mK and cooling to
1.2$\pm$0.1 mK (extracted using the method described above), demonstrating a reduction in temperature upon demagnetization by a factor
of ten. The other chips perform similar to D, thus substantiating the overall cooling scheme proposed here for reaching submillikelvin temperatures on multiple measurement leads.

Ideally, $T_i$ of the Cu nuclei will be reduced by the same factor $x$ as the $B$-field. To characterize the demagnetization process, we introduce
the efficiency $\xi_{B_f}=T_i/T_f \div B_i/B_f$, and find $\xi_{4 T}=88 \pm 3$\%, $\xi_{2 T}=80 \pm 3$\%, $\xi_{1 T}=63 \pm 3$\%, and $\xi_{0.3 T}=42
\pm 2$\%. $\xi<100$\% signifies nonadiabadicity, presumably due to a spurious heat leak. The lower efficiency at smaller $B_f$ probably reflects the
reduction in stored energy of the nuclei. The dominant loss mechanism is presently unclear and under investigation. Parasitic coupling between NRs
has been seen from power curves, thus emphasizing one of the many technical challenges of building a parallel NR network.

In conclusion, we have laid out a method that enables the direct and simultaneous cooling of the electrical leads to a nanoscale device. Its strength
is that it short circuits the two main bottlenecks of cooling electrons: thermal boundary resistance and electron-phonon coupling. We have addressed
the technical challenges of constructing a parallel NR network, yielding a prototype that achieves a base temperature of $\sim1\,$mK. Future efforts will reduce the intrinsic heat leak and address the presently low efficiency (securing temperatures well within the microkelvin regime), further demonstrate ultralow temperatures directly in nanoscale devices, and add an independently controllable magnetic field for the sample.

We thank G. R. Pickett, R. P. Haley, M. Paalanen, R. Blaauwgeers, G. Frossati, and A. de Waard for their advice. This work was supported by the Swiss
Nanoscience Institute, Swiss NSF, ERC Starting Grant, and EU-FP7 MICROKELVIN network.


\providecommand{\noopsort}[1]{}\providecommand{\singleletter}[1]{#1}%

\end{document}